\documentclass[sigconf]{acmart}

\newcommand{\method}{ProxyWar\xspace}
\usepackage{xspace}
\usepackage{enumitem}
\usepackage{multirow}
\usepackage{colortbl}
\usepackage{subcaption}
\usepackage{listings}
\lstdefinestyle{proxywarpython}{
    language=Python,
    basicstyle=\ttfamily\small,
    keywordstyle=\color{blue!80!black}\bfseries,
    commentstyle=\color{gray!70!black},
    stringstyle=\color{orange!90!black},
    numberstyle=\tiny\color{gray!80},
    numbers=left,
    numbersep=6pt,
    xleftmargin=2em,
    frame=single,
    rulecolor=\color{gray!40!white},
    backgroundcolor=\color{white},
    showstringspaces=false,
    breaklines=true,
    tabsize=4,
    columns=fullflexible,
    captionpos=b,
    aboveskip=1em,
    belowskip=1em,
    emph={BaseAgent,ABC,self}, emphstyle={\color{teal}\bfseries},
    emph={[2]str,int,bool,Any,List,Optional}, emphstyle={[2]\color{violet}},
    emph={[3]@abstractmethod}, emphstyle={[3]\color{red!70!black}\bfseries}, 
}

\newcounter{finding}
\newenvironment{finding}{%
  \refstepcounter{finding}%
  \par\noindent
  \setlength{\tabcolsep}{6pt}%
  \renewcommand{\arraystretch}{1.3}%
  \begin{center}
  \begin{tabular}{@{}p{\linewidth}@{}}
    \rowcolor{gray!10}%
    \textbf{Finding~\thefinding:}\ \itshape
}{%
  \end{tabular}%
  \end{center}
  \par
}

\AtBeginDocument{%
  }

\copyrightyear{2026}
\acmYear{2026}
\setcopyright{cc}
\setcctype{by}
\acmConference[ICSE '26]{2026 IEEE/ACM 48th International Conference on Software Engineering}{April 12--18, 2026}{Rio de Janeiro, Brazil}
\acmBooktitle{2026 IEEE/ACM 48th International Conference on Software Engineering (ICSE '26), April 12--18, 2026, Rio de Janeiro, Brazil}
\acmDOI{10.1145/3744916.3773220}
\acmISBN{979-8-4007-2025-3/26/04}

\begin{document}
\title{ProxyWar: Dynamic Assessment of LLM Code Generation in Game Arenas}

\author{Wenjun Peng}
\email{wenjun.peng@adelaide.edu.au}
\authornote{These authors contributed equally to this work.}
\author{Xinyu Wang}
\email{xinyu.wang02@adelaide.edu.au}
\authornotemark[1]
\authornote{Xinyu Wang is the corresponding author.}
\author{Qi Wu}
\email{qi.wu01@adelaide.edu.au}

\begin{abstract}
Large language models (LLMs) have revolutionized automated code generation, yet the evaluation of their real-world effectiveness remains limited by static benchmarks and simplistic metrics. We present \method, a novel framework that systematically assesses code generation quality by embedding LLM-generated agents within diverse, competitive game environments. Unlike existing approaches, \method evaluates not only functional correctness but also the operational characteristics of generated programs, combining automated testing, iterative code repair, and multi-agent tournaments to provide a holistic view of program behavior. Applied to a range of state-of-the-art coders and games, our approach uncovers notable discrepancies between benchmark scores and actual performance in dynamic settings, revealing overlooked limitations and opportunities for improvement. These findings highlight the need for richer, competition-based evaluation of code generation. Looking forward, \method lays a foundation for research into LLM-driven algorithm discovery, adaptive problem solving, and the study of practical efficiency and robustness, including the potential for models to outperform hand-crafted agents. The project is available at \url{https://github.com/xinke-wang/ProxyWar}.
\end{abstract}



\maketitle

\section{Introduction}

The emergence of Large Language Models (LLMs), such as GPT-4~\cite{achiam2023gpt} and AlphaCode~\cite{li2022competition}, has fundamentally transformed the landscape of software development~\cite{fan2023large}. Unlike traditional code generation tools that rely on fixed templates or rule-based logic~\cite{bruch2009learning, hindle2016naturalness}, LLMs are capable of producing diverse, context-aware, and often highly creative code snippets directly from natural language instructions~\cite{chen2024survey, gao2025current}. With their remarkable ability to understand complex programming tasks and synthesize functioning programs across multiple languages and domains~\cite{fan2023automated, li2022competition}, LLM-based code generation tools are increasingly integrated into modern development workflows. This shift is not only accelerating the pace of software engineering, but also reshaping the operational workflows of programmers, including problem-solving, debugging, code review, and system design~\cite{gao2025current}.

\begin{figure}[t!]
    \centering
    \includegraphics[width=\linewidth]{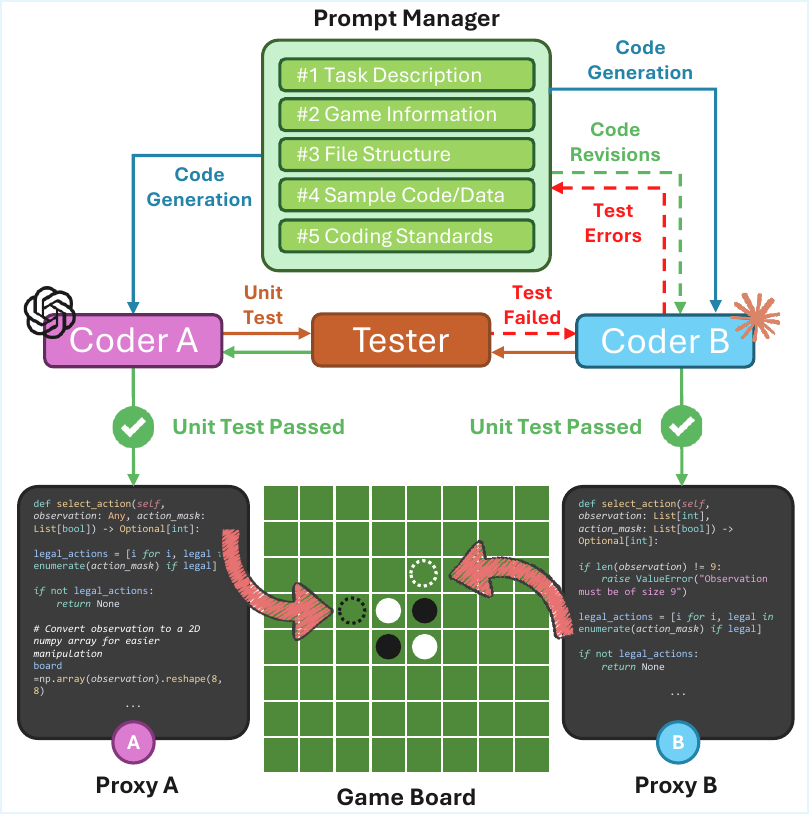}
    \caption{\textbf{Overview of the \method\ framework.} The Prompt Manager provides standardized game specifications to multiple LLM code generators, which develop game-playing agents. The Tester validates each agent through unit tests, with failed tests triggering a repair loop that sends error messages back for revision. Successfully tested agents compete as proxies on the Game Board in automated tournaments. Match outcomes determine skill ratings and provide a comprehensive evaluation of code generation quality through both functional correctness and competitive performance.}
    \Description{System overview diagram illustrating the ProxyWar framework. The Prompt Manager produces standardized game specifications that are sent to multiple LLM-based code generators to create game-playing agents. A Tester component evaluates each agent using unit tests and returns error messages to trigger an iterative repair loop when tests fail. Agents that pass testing are evaluated on a Game Board, where they compete in automated tournaments to generate match outcomes and skill ratings.}
    \label{fig:overview}
\end{figure}

However, despite the rapid adoption of LLM-based code generation, the methods for evaluating these models remain largely inadequate~\cite{liu2023your}. Current approaches~\cite{chen2021evaluating, chen2024survey, yu2024codereval, hendrycks2021measuring, agarwal2024copilot, dibia2022aligning} predominantly rely on metrics borrowed from natural language processing (e.g., BLEU scores~\cite{papineni2002bleu}) or simplistic functional tests (e.g., pass@k on HumanEval~\cite{chen2021evaluating}). While these metrics provide some insights into syntactic similarity or basic correctness, they fail to capture the multifaceted nature of operational characteristics of generated code (e.g., runtime stability, efficiency, and robustness) that matter in real-world software development~\cite{ren2020codebleu}. A model that generates code passing all test cases might produce solutions that are inefficient, unmaintainable, or brittle when faced with edge cases~\cite{chen2024survey}. More critically, many widely used benchmarks evaluate code in isolation, largely emphasizing function-level correctness without capturing behavior under dynamic execution constraints, even though some recent efforts (e.g., agent-based SWE-bench~\cite{jimenez2024swebench} evaluations) begin to introduce limited interaction. This leaves open the broader question of \textit{how to evaluate systems where code must not only work but also operate reliably and competitively under shared resource budgets}.

This limitation becomes particularly apparent when considering domains like algorithmic competitions, game AI development, performance optimization, or self-improvement workflows, where the usefulness of code is naturally measured by its comparative performance against alternative solutions. For example, in competitive programming, solutions are evaluated on hidden test cases, time/memory limits, and runtime performance, meaning that two functionally correct programs can differ substantially in their practical effectiveness~\cite{li2022competition}. In these contexts, traditional pass/fail metrics provide little insight into whether one model produces code that is more efficient, more stable, or more strategically effective than another. The lack of comprehensive, comparative evaluation frameworks has created a significant gap between the promise of LLM-based code generation and our ability to systematically assess their real-world viability.

To address this challenge, we present \textbf{\method}, a novel framework that evaluates LLM-generated code through competitive, controlled execution under shared resource budgets. Instead of focusing solely on static correctness, \method embeds generated agents into dynamic game environments, where head-to-head competition provides a practical and fine-grained signal of a model’s ability to produce code that is not only correct but also efficient, stable, and adaptable. As illustrated in Figure~\ref{fig:overview}, \method\ orchestrates a complete pipeline from code generation to competitive evaluation: LLMs first generate game-playing agents based on standardized prompts; these agents undergo hierarchical unit testing with an optional repair loop; and finally they compete in automated tournaments where their performance provides a holistic measure of operational behavior rather than merely functional success.

Unlike static benchmarks, \method\ evaluates code in dynamic, interactive environments where success depends not just on correctness but also on algorithmic efficiency, runtime robustness, adaptability, and decision-making under constraints. The framework supports multiple code generators competing simultaneously, enabling comprehensive comparisons between different models or configurations. The integrated testing and revision mechanism ensures that we measure not only raw generation capability but also the ability of a model to debug, refine, and stabilize its own code, a property increasingly important in autonomous software engineering workflows. Through automated tournaments and skill-based rankings using algorithms like TrueSkill~\cite{minka2018trueskill}, \method provides a flexible approach to compare different code generation models across multiple dimensions of code quality. This paper makes the following contributions:

\begin{itemize}[leftmargin=2em]
    \item \method, the first competitive, execution-based evaluation framework for LLM code generation, which orchestrates automated execution, testing, and iterative repair mechanisms to evaluate code quality through competitive gameplay, supporting three key capabilities:
    \begin{enumerate}[leftmargin=2.5em]
        \item It automatically generates, tests, and deploys game-playing agents from natural language specifications.
        \item It facilitates iterative code improvement through automated error feedback and repair loops.
        \item It provides skill-based rankings that reflect both functional correctness and operational performance.
    \end{enumerate}
    \item A multi-dimensional assessment methodology that captures diverse aspects of operational characteristics beyond traditional pass/fail metrics, including algorithmic efficiency, code complexity, runtime performance, and strategic decision-making capabilities.
    \item Implementation of a scalable platform supporting diverse game environments, from perfect information board games to imperfect information card games, enabling systematic assessment across varied programming challenges.
    \item Extensive empirical evaluation on state-of-the-art LLMs including ChatGPT, Claude, Gemini, and leading open-source models. By combining traditional metrics with competitive performance analysis, our results reveal previously hidden limitations in LLMs' code generation capabilities, particularly in algorithmic creativity, optimization strategies, and adaptive problem-solving, providing new insights into the gap between functional correctness and practically deployable behavior.
\end{itemize}

\section{Background and Motivation}

Code generation has evolved significantly from early template-based and pattern-based approaches to recent LLM-based methods. Prior to the LLM era, code generation systems relied on various techniques including API pattern mining~\cite{zhong2009mapo, wang2013mining}, statistical language models~\cite{hindle2016naturalness, allamanis2013mining}, and neural approaches like sequence-to-sequence models~\cite{ling2016latent, yin2017syntactic}. However, the evaluation methods for these diverse code generation systems have not kept pace with their rapid advancement. While traditional generators could be evaluated through API coverage metrics and simple correctness checks, the emergence of LLMs capable of generating diverse, creative solutions has exposed fundamental limitations in existing evaluation methodologies. In this section, we examine the evolution of code generation evaluation, analyze the limitations of current approaches, and establish the motivation for our game-based evaluation framework. This paper primarily focuses on evaluating LLM-based code generation, as these models now dominate both research and practical applications in automated programming.

\subsection{Code Generation Quality Assessment}

The journey from text-based to execution-based evaluation represents a fundamental paradigm shift in how we assess generated code. Early evaluation approaches borrowed heavily from natural language processing, treating code as sequences of tokens rather than executable logic.

\noindent\textbf{BLEU and Text-based Metrics.} The BLEU score~\cite{papineni2002bleu}, originally designed for machine translation, measures n-gram overlap between generated and reference code: $\text{BLEU} = \text{BP} \cdot \exp\left( \sum_{n=1}^N w_n \cdot \log(p_n) \right)$, where $p_n$ represents the precision of n-grams and BP is a brevity penalty. However, BLEU treats all tokens equally and ignores code structure, so two functionally equivalent implementations with different variable names or loop structures receive low scores despite identical behavior~\cite{ren2020codebleu}.

\noindent\textbf{CodeBLEU: Incorporating Code Structure.} Recognizing the limitations of BLEU, CodeBLEU~\cite{ren2020codebleu} extends BLEU by incorporating syntactic and semantic features: $\text{CodeBLEU} = \alpha \cdot \text{BLEU} + \beta \cdot \text{BLEU}_\text{weight} + \gamma \cdot \text{Match}_\text{AST} + \delta \cdot \text{Match}_\text{DF}$. This metric combines weighted n-gram matching (with keywords weighted 5× higher), Abstract Syntax Tree (AST) similarity, and data-flow analysis. While CodeBLEU achieves higher correlation with human evaluation, it still relies on reference solutions and cannot assess functional correctness~\cite{ren2020codebleu}.

\noindent\textbf{Pass@k: The Shift to Execution-Based Evaluation.} The introduction of HumanEval~\cite{chen2021evaluating} marked a pivotal transition to execution-based metrics. The pass@k metric measures the probability that at least one of k generated samples passes all test cases: $\text{pass@}k = 1 - \frac{\dbinom{n-c}{k}}{\dbinom{n}{k}}$, where $n$ is the total number of samples, $c$ is the number of correct samples, and $k$ is the number of samples considered. This metric directly evaluates functional correctness but introduces new challenges around test adequacy and coverage.

\subsection{Limitations of Current Approaches}

Despite advances in metrics, fundamental limitations persist in how we evaluate LLM-generated code. These limitations span multiple dimensions, from dataset quality to evaluation scope.

\noindent\textbf{Simplicity Bias and Limited Scope.} Current benchmarks exhibit severe simplicity bias. HumanEval~\cite{chen2021evaluating} contains only 164 problems, averaging 7.7 test cases each, focusing on algorithmic puzzles rather than real programming tasks. Analysis of real codebases reveals that over 70\% of functions are non-standalone, requiring understanding of complex dependencies and cross-file relationships~\cite{du2024evaluating}. Yet mainstream benchmarks evaluate only isolated functions. The MBPP benchmark~\cite{austin2021program}, while larger at 974 problems, suffers from similar limitations with mostly single-function tasks under 50 lines. Recent work, such as SWE-bench~\cite{jimenez2024swebench} and ClassEval~\cite{du2024evaluating} partially mitigates simplicity bias by introducing multi-file or multi-class settings, but these benchmarks still evaluate code in a non-interactive manner and do not incorporate execution-time resource constraints.

\noindent\textbf{Dataset Contamination and Memorization.} Recent studies have uncovered widespread dataset contamination across benchmarks~\cite{jimenez2024swebench}. Analysis found that 94\% of SWE-bench issues were created before LLM training cutoffs~\cite{aleithan2024swe}, with clear evidence of memorization affecting results~\cite{chen2025memorize}. When evaluated on truly held-out test sets, model performance drops significantly, for example, StarCoder-7B~\cite{lozhkov2024starcoder} achieved Pass@1 scores 4.9 times higher on leaked samples than non-leaked samples~\cite{zhou2025lessleak}. While game environments are also publicly available and therefore not immune to contamination, competitive settings reduce the effectiveness of memorized solutions: agents must handle diverse opponent behaviors, adhere strictly to environment interfaces, and operate under runtime constraints, making brittle or overfitted memorized implementations more likely to fail.

\noindent\textbf{Static Evaluation Misses Dynamic Reality.} Perhaps most critically, current evaluation is fundamentally static while real programming is inherently dynamic. Professional development involves iterative refinement, debugging based on compiler feedback, and collaborative problem-solving, but these dynamic aspects are largely absent from existing benchmarks~\cite{zhang2024codeagent}. Evaluating models solely on isolated functions fails to capture their effectiveness in realistic, end-to-end software engineering workflows. Meanwhile, recent agent-based evaluations (e.g., SWE-bench agents, CodeAgent~\cite{zhang2024codeagent}) introduce limited forms of interactivity, but they typically operate under single-agent, non-adversarial settings and do not compare multiple code generators under shared constraints. Similar challenges have been noted even for general-purpose LLMs, where static or narrowly scoped benchmarks often fail to reflect real-world reasoning, perception, or interaction capabilities~\cite{qiao2025navbench, wang2024modaverse, wang2020general, fu2024ocrbench}. These observations parallel our setting: static correctness tests in code generation likewise miss the dynamic, context-dependent behaviors required for practical performance.

\noindent\textbf{Limited Quality Dimensions.} Existing metrics focus almost exclusively on functional correctness, ignoring other crucial aspects of operational characteristics, such as:

\begin{itemize}[leftmargin=*]
    \item \textbf{Efficiency}: Algorithmic complexity or runtime performance.
    \item \textbf{Robustness}: Sensitivity to input variations and edge cases remains largely untested.
    \item \textbf{Maintainability}: Code readability, documentation, and modular design are not assessed.
    \item \textbf{Creativity}: Novel algorithmic approaches versus memorized solutions cannot be distinguished.
\end{itemize}

\begin{figure*}[t!]
    \centering
    \includegraphics[width=0.8\linewidth]{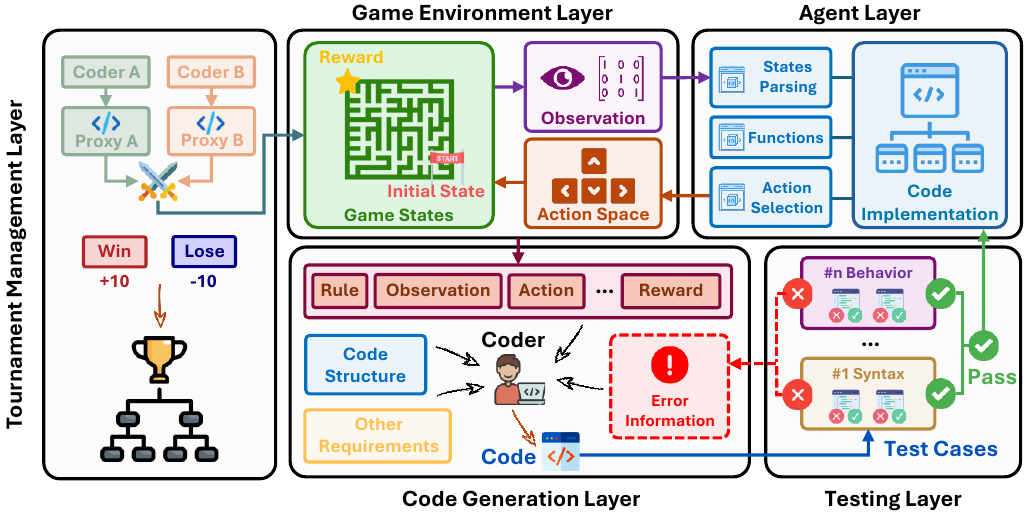}
    \caption{\textbf{The ProxyWar framework pipeline.} ProxyWar evaluates LLM code generation through a multi-layer architecture. The process begins with the Code Generation Layer, where LLMs (Coders) receive game specifications (rules, observation/action formats, etc.) to generate agent implementations. Failed attempts trigger an iterative repair loop with detailed error feedback. The Testing Layer validates generated code through hierarchical test cases. Successfully tested code is deployed in the Agent Layer, which parses game states and executes action selection logic. Finally, the Tournament Management Layer orchestrates competitions between agents (Proxy A vs. Proxy B) in the Game Environment, where agents receive observations, select actions, and compete for rewards. Match outcomes are aggregated into rankings, providing a comprehensive assessment of code generation quality based on actual competitive performance rather than static metrics.}
    \Description{Pipeline diagram of the ProxyWar framework showing a multi-layer evaluation process for LLM-generated code. The pipeline starts with a Code Generation Layer where LLM-based coders generate agent implementations from game specifications and receive iterative error feedback for repair. A Testing Layer validates the generated code using hierarchical test cases. Code that passes testing is deployed in an Agent Layer that interprets game states and selects actions. A Tournament Management Layer then coordinates competitions between agents in a game environment, aggregates match outcomes, and produces rankings that reflect competitive performance.}
    \label{fig:pipeline}
\end{figure*}

\subsection{The Gap Between Benchmarks and Real-World Programming}

The disconnect between benchmark performance and practical utility has become increasingly apparent. GitHub Copilot studies~\cite{ziegler2024measuring} show that user acceptance rate, not correctness metrics, best predicts productivity gains. This suggests that current evaluation frameworks fail to capture what makes code genuinely useful.

\noindent\textbf{Missing Contextual Understanding.} Real-world programming tasks demand an understanding of existing codebases, API constraints, and architectural decisions. ClassEval~\cite{du2024evaluating} demonstrates that class-level code generation is significantly more challenging than function-level generation, with even the most capable LLMs exhibiting notably low success rates when dependencies are involved. This lack of contextual understanding suggests that high benchmark scores may not accurately reflect practical ability.

\noindent\textbf{Absence of Interactive Development.} Professional programmers spend most of their time debugging and refining code rather than writing it from scratch~\cite{beller2015much}. However, few existing benchmarks assess iterative improvement based on test failures or runtime behavior. The recent CodeAgent framework~\cite{zhang2024codeagent} showed that allowing models to iterate based on feedback significantly improves success rates, highlighting the importance of interactive evaluation. Collectively, these works demonstrate increasing interest in interactive and agent-based evaluation. However, they do not provide a comparative, head-to-head assessment of generated programs under shared resource constraints, nor do they measure operational robustness in dynamic environments.

\noindent\textbf{Limited Competitive Assessment.} In many domains, such as algorithmic competitions, game AI, and optimization tasks, code quality is inherently comparative. A solution that merely works may be vastly inferior to one that works efficiently. Current pass/fail metrics cannot capture these performance gradients or trade-offs between different quality dimensions.

\subsection{Motivation}

The limitations of existing evaluation approaches motivate our game-based framework. Games provide unique advantages for assessing code generation capabilities:

\noindent\textbf{Natural Performance Gradients.} Unlike binary pass/fail tests, game competitions produce continuous performance metrics through score differentials. This enables fine-grained comparison between models that would appear equivalent under traditional metrics. For example, a chess engine that wins 60\% of games clearly outperforms one winning 40\%, even if both implement legal moves correctly.

\noindent\textbf{Reduced Reliance on Memorization.} Because the space of possible game states is large and runtime constraints limit exhaustive search, game environments reduce the effectiveness of purely memorized or template-based approaches and make brittle, overfitted strategies more likely to fail.

\noindent\textbf{Multi-Dimensional Assessment.} Game environments naturally evaluate multiple quality dimensions simultaneously:

\begin{itemize}[leftmargin=*]
    \item \textbf{Correctness}: Invalid moves are immediately apparent.
    \item \textbf{Efficiency}: Slow algorithms lose on time or get outmaneuvered.
    \item \textbf{Robustness}: Strategies must handle diverse opponent behaviors.
    \item \textbf{Adaptability}: Success requires adapting to situations.
\end{itemize}

This aligns naturally with evaluating the operational characteristics of generated code rather than only its syntactic or functional correctness.

\noindent\textbf{Scalable Complexity.} Games provide a controllable range of complexity, from simple ones like tic-tac-toe to more complex ones like chess, allowing assessment across different skill levels. Even simple games serve as useful calibration environments for testing legality, basic decision logic, and robustness under repeated interaction.

By situating code generation evaluation within competitive game environments, we can assess not just whether models can write code, but whether they can write \textit{good} code that performs well under challenging, dynamic conditions. This approach addresses the critical gaps in existing evaluation frameworks while providing actionable insights into model capabilities and limitations.

\section{The ProxyWar Framework}

\subsection{Design Principles}

The \method framework is built upon four fundamental design principles that guide its architecture and implementation:

\noindent\textbf{P1: Competitive Evaluation through Direct Competition.} Unlike static benchmarks that evaluate code in isolation, \method places generated agents in direct competition. This principle ensures that the operational characteristics of generated programs are measured not by adherence to predetermined patterns but by their actual performance under shared constraints and against diverse opponents. The competitive nature reveals subtle quality differences that binary pass/fail tests cannot capture.

\noindent\textbf{P2: Separation of Concerns between Framework and Agent.} \method adopts a strict separation where the framework handles all game mechanics, state management, and rule enforcement, while agents focus on decision-making. This design choice simplifies the code generation task and ensures fair comparison; all agents operate under identical conditions with the same computational constraints. Crucially, this separation treats the environment as a pluggable task sandbox, so the same protocol can, in principle, be applied to non-game software engineering tasks (e.g., bug fixing or performance tuning) that can be executed in a controlled environment.

\noindent\textbf{P3: Multi-Dimensional Quality Assessment.} \method targets multiple dimensions of behavior beyond functional correctness, including algorithmic efficiency, robustness to edge cases, runtime stability, and strategic decision-making. Each dimension contributes to overall performance but can be analyzed independently.

\noindent\textbf{P4: Iterative Refinement through Automated Feedback.} Real-world programming is iterative. \method incorporates automated testing and repair loops, allowing generated code to be refined based on error feedback. This mirrors professional development practices and evaluates not just initial outputs but also the model's debugging capabilities.

\subsection{System Architecture}

As demonstrated in Figure~\ref{fig:pipeline}, \method comprises five primary layers, each with well-defined responsibilities and interfaces.

\noindent\textbf{Game Environment Layer} provides a unified interface. Formally, a game environment can be defined as a tuple: $\mathcal{G} = \langle \mathcal{S}, \mathcal{A}, \mathcal{T}, \mathcal{R}, \mathcal{O}, s_0 \rangle$. In this formulation, $\mathcal{S}$ denotes the state space and $\mathcal{A}$ the action space. The transition function $\mathcal{T}: S \times A \to \mathbb{R}$ assigns a scalar reward to each state-action pair. The observation function $\mathcal{O}: S \to \mathcal{O}_{\text{space}}$ maps the underlying state to the observation available to an agent, accommodating both perfect information ($\mathcal{O}(s) = s$) and imperfect information scenarios. The initial state $s_{0} \in S$ specifies where the environment begins. This formalization supports a wide range of games with different information structures, while the environment layer enforces game rules, manages state transitions, and ensures consistent observations for all agents. In non-game settings, the same abstraction can represent other executable SE tasks, where \textit{states} are program or system configurations and \textit{actions} are candidate edits or decisions.

\noindent\textbf{Agent Layer} defines the minimal interface that all generated code must implement. $\pi$ is the policy function: $\pi : \mathcal{O}_{\text{space}} \times \{0, 1\}^{|\mathcal{A}|} \to \mathcal{A}$, where the second parameter serves as an action mask indicating legal moves. Additional methods and attributes can be freely added, while this interface is required to ensure compatibility. This minimal contract is deliberately simple so that diverse models and coding styles can be evaluated under a common protocol.

\begin{table*}[t!]
\centering
\caption{Characteristics of the game environments in \method. State-space sizes are approximate legal-position counts; action-space sizes are the maximum number of legal actions from any state.}
\label{tab:games}
\begin{tabular}{lcccc}
\toprule
\textbf{Game} & \textbf{State Space} & \textbf{Action Space} & \textbf{Information} & \textbf{Complexity} \\
\midrule
\multicolumn{5}{l}{\textit{Single-Player Puzzles}} \\
Sudoku & $6.67\times10^{21}$ & $\le 9$ / cell & Perfect & NP-complete \\
2048 (4$\times$4) & $\sim4.4\times10^{16}$ & $4$ & Perfect+Random & NP-hard \\
Tower of Hanoi ($n$) & $3^{n}$ & $\le 6$ & Perfect & $O(2^{n})$ optimal \\
Maze (grid) & $W \times H$ & $\le 4$ & Perfect / Partial & $O(V+E)$ \\
\midrule
\multicolumn{5}{l}{\textit{Two-Player Board/Spatial Games}} \\
Tic-Tac-Toe & $1.97\times10^{4}$ & $\le 9$ & Perfect & Solved / $O(1)$ \\
Connect Four & $4.53\times10^{12}$ & $\le 7$ & Perfect & PSPACE-complete \\
Reversi & $\sim10^{28}$ & $\sim 10$ & Perfect & PSPACE-complete \\
Snake (2-player, $n{\times}n$) & $2^{n^{2}}$ & $4$ & Perfect & PSPACE-complete \\
\midrule
\multicolumn{5}{l}{\textit{Multi-Player Card Games}} \\
Texas Hold’em (Limit) & $\sim1.6\times10^{17}$ & $\le 4$ & Imperfect & PSPACE-complete \\
\bottomrule
\end{tabular}
\end{table*}

\noindent\textbf{Code Generation Layer} manages the transformation from natural language specifications to executable agent code. For each game $\mathcal{G}$ and model $M$, the Generation Process can be formulated as: $(M, \text{Prompt}(\mathcal{G})) \rightarrow \text{Code}_\pi$. This layer also provides a Repair Function for bug fixing: $(M, \text{Code}_\pi, \text{Errors}) \rightarrow \text{Code}_{\pi'}$, which enables multi-round improvement based on test feedback, effectively capturing the model’s debugging capabilities. The same interface is used regardless of whether the task is a game, a regression bug fix, or a performance optimization problem, making the pipeline reusable across different SE settings.

\noindent\textbf{Testing Layer} is responsible for validating the generated agent code through a hierarchical testing strategy. This layer systematically checks each agent using a test suite $\mathcal{T} = \{t_1, t_2, ..., t_n\}$, which covers multiple levels of validation: syntax checking, import verification, interface compliance, and behavioral correctness. Each level is designed to detect different types of errors, ranging from basic syntax issues to violations of required interfaces or incorrect game logic. Formally, let $\mathcal{T}$ be the set of all test cases. The testing function is defined as follows:

\[
    \texttt{Test}(\text{Code}_\pi, \mathcal{T}) \rightarrow 
    \begin{cases}
    \text{PASS} & \text{if } \forall t_i \in \mathcal{T}: t_i(\text{Code}_\pi) = \text{true} \\
    (\text{FAIL}, E) & \text{otherwise, where } E \text{ is the error set}
    \end{cases}
\]

This layer enables early detection of errors, ensuring that only code meeting all requirements proceeds to further evaluation or competition. In practice, \method combines generic tests (e.g., AST and interface checks that apply to any environment) with environment-specific suites (e.g., multi-move interactions or edge-case scenarios), all executed by a common harness without human intervention at evaluation time. In addition, any detected errors and their details are provided to the Code Generation Layer, enabling the repair function to iteratively refine the agent code based on concrete feedback.

\noindent\textbf{Tournament Management Layer.} The Tournament Management Layer orchestrates competitive evaluation by scheduling matches among agents. While a full round-robin format is employed for small numbers of agents or two-player games, the framework also supports sampling-based tournament scheduling for scenarios with many agents or multiplayer games. In these cases, a representative subset of matches is sampled to ensure metric convergence without incurring excessive computational cost. Formally, the tournament can be represented as a directed graph $G = (V, E)$, where $V = \{\pi_1, \pi_2, \ldots, \pi_n\}$ denotes the set of agents, and $E = \{(i, j, w) \mid i, j \in V,\, w \in \{i, j, \text{draw}\}\}$ captures the observed outcomes of the sampled matches, with $w$ indicating the winner or a draw. This layer provides the flexibility to support both exhaustive and sampled tournaments, accommodating different game types and evaluation budgets while ensuring reliable agent ranking and metric stability. Because the scheduling logic is environment-agnostic, the same engine can be reused for head-to-head evaluation on non-game SE tasks (e.g., comparing two bug-fixing agents on the same issue set under identical resource limits).

\subsection{Multi-Dimensional Evaluation Methodology}

\method evaluates generated code across multiple dimensions, providing a comprehensive assessment beyond simple win rates.

\noindent\textbf{Competitive Performance Metrics.} We employ the TrueSkill rating system~\cite{minka2018trueskill} proposed by Microsoft Xbox Network to compute skill ratings. Each agent $i$ is modeled by a skill distribution: $\text{skill}_i \sim \mathcal{N}(\mu_i, \sigma_i^2)$. After each match, the ratings are updated via Bayesian inference. To ensure robust comparisons, we report the conservative skill estimate defined as $\mu_i - 3\sigma_i$, which provides a lower bound on expected performance with 99.7\% confidence. This framework is applied consistently across both multiplayer and single-player games. In the multiplayer setting, TrueSkill is used in the standard manner to update skills based on the outcomes among all participants. In the case of single-player games, agents are evaluated by performing their tasks independently under the same conditions, and their results are compared to determine relative performance. These comparisons are then incorporated into the TrueSkill system in the same way as multiplayer outcomes. This unified methodology enables fair and consistent ranking of agents across all game types in \method.

\noindent\textbf{Static Evaluation Metrics.} Beyond competitive outcomes, \method\ evaluates static code quality using a suite of hierarchical, task-driven metrics tailored to code generation scenarios. These metrics are designed to complement tournament-based evaluation by revealing properties that are not always reflected in match results:

\begin{itemize}[leftmargin=*]
    \item \textbf{Pass@1}: The proportion of initial agent submissions that pass all reference test cases, capturing basic functional correctness.
    \item \textbf{Repair Rate}: The rate at which agents successfully fix initial bugs and achieve a passing solution within a limited number of iterations, measuring self-debugging ability.
    \item \textbf{Hierarchical Test Layers}:
    \begin{itemize}
        \item \textbf{Structure}: Syntactic correctness and conformity to required interfaces.
        \item \textbf{Function}: Ability to solve fundamental subtasks or simple test cases.
        \item \textbf{Logic}: Success on more complex, game-specific or edge-case logic tests.
        \item \textbf{Robustness}: Stability and correctness across adversarial or stress test scenarios.
    \end{itemize}
\end{itemize}

Together, these metrics characterize the operational behavior of each generated agent in terms of its activity frequency, repair reliability, and stress tolerance, independent of its relative ranking in tournaments. By combining these static metrics with dynamic tournament results, \method enables comprehensive, multi-dimensional assessment of LLM-generated code.

\section{Implementation}

This section details the implementation of \method, highlighting the engineering challenges and solutions involved in building a scalable, game-based code evaluation system. Our entire evaluation pipeline is based on \textit{\textbf{Python}} code generation, a deliberate choice for its simplicity, ecosystem maturity, and strong support among LLMs. Python's wide adoption in both research and industry facilitates reproducibility and enables seamless integration with diverse game environments, agent implementations, and analysis tools. In this section, we discuss game environment design, coder integration, prompt engineering, testing infrastructure, and practical considerations for extensibility and reliability.

\begin{table}[t]
\centering
\caption{LLM-based coders integrated in \method. ``O/S” indicates open source status. Context is the maximum token window.}
\label{tab:coders}
\begin{tabular}{@{\hspace{2pt}}l@{\hspace{2pt}}l@{\hspace{2pt}}c@{\hspace{2pt}}c@{\hspace{2pt}}r@{\hspace{2pt}}}
\toprule
\textbf{Release} & \textbf{Model} & \textbf{Size} & \textbf{O/S} & \textbf{Context} \\
\midrule
\multicolumn{5}{l}{\textit{General-purpose}} \\
2024.9 & Qwen2.5-72B~\cite{bai2025qwen2} & 72 B & Y & 32,768 \\
2024.10& Claude3.5-Sonnet~\cite{anthropic2024claude35} & — & N & 200,000 \\
2025.1 & Phi4~\cite{abdin2024phi} & 14 B & Y & 16,384 \\
2025.2 & Gemini2.0-Flash~\cite{google2024gemini2} & — & N & 1,048,576 \\
2025.3 & DeepSeekV3-0324~\cite{liu2024deepseek} & 685 B & Y & 163,840 \\
2025.4 & Llama4-Maverick~\cite{meta2024llama4}  & 400 B & Y & 1,048,576 \\
2025.4 & GPT4.1-Mini-20250414~\cite{openai2024gpt41} & — & N & 1,047,576 \\
2025.4 & GPT4.1-20250414~\cite{openai2024gpt41} & — & N & 1,047,576 \\
\midrule
\multicolumn{5}{l}{\textit{Reasoning-enhanced}} \\
2025.1 & O3-Mini-20250131~\cite{openai2025o3mini} & — & N & 200,000 \\
2025.4 & Qwen3-235B-0428~\cite{yang2025qwen3} & 235 B  & Y & 40,960 \\
2025.5 & DeepSeek-R1-0528~\cite{guo2025deepseek} & 671 B  & Y & 128,000 \\
2025.5 & Claude4-Sonnet-20250522~\cite{anthropic2025claude4} & — & N & 200,000 \\
2025.6 & Gemini-2.5-Flash~\cite{geminiteam2025gemini25} & — & N & 1,048,576 \\
2025.6 & Magistral-Small-2506~\cite{rastogi2025magistral} & 24 B & Y & 40,000 \\
\midrule
\multicolumn{5}{l}{\textit{Code-specialized}} \\
2024.11& Qwen2.5-Coder~\cite{hui2024qwen2} & 32 B  & Y & 32,768 \\
2025.1 & Codestral-2501~\cite{mistral2024codestral} & — & N & 262,144 \\
2025.4 & Mercury-Coder~\cite{labs2025mercury} & — & N & 32,000 \\
2025.5 & Codex-Mini~\cite{openai2025codexmini} & — & N & 200,000 \\
\bottomrule
\end{tabular}
\end{table}

\subsection{Game Environments}

\method implements ten carefully selected games spanning three categories: single-player puzzles, two-player board games, and multiplayer card games. This diversity ensures comprehensive evaluation across a wide range of algorithmic paradigms, information structures, and programming challenges. Each game is chosen to stress different aspects of code generation, such as search, planning, probabilistic reasoning, and collaboration under uncertainty.

Table~\ref{tab:games} summarizes the core characteristics of the implemented game environments, including state space size, action space, information structure, and computational complexity. The selected games collectively cover a broad spectrum of complexity and uncertainty, from fully solved, perfect-information puzzles to challenging multiplayer games with hidden information and stochastic dynamics. This diversity ensures that \method can evaluate code generation models on a wide range of algorithmic skills, including planning, search, probabilistic reasoning, and adaptive decision-making under uncertainty. By integrating both classic and modern programming challenges, \method provides a robust foundation for comprehensive evaluation of code generation models.

\subsection{Coder Integration}

\method is designed to enable flexible evaluation of code generation systems. While this paper focuses on LLM-based coders, reflecting their prevalence in current research and deployment, the framework itself adopts a modular interface that allows for straightforward integration of new coders, regardless of the architecture.

To streamline integration and experiment management, \method leverages the OpenRouter~\footnote{https://openrouter.ai/} platform, which aggregates APIs from major providers, including OpenAI, Anthropic, Google, and Meta, behind a single endpoint. This design minimizes engineering overhead, supports unified monitoring and analytics, and allows new coders to be added or replaced through simple configuration changes without modifying the core framework code.

Table~\ref{tab:coders} provides an overview of the coders evaluated in this work. The selection includes both proprietary and open-source models and covers a range of parameter scales and context window lengths. For analysis, we organize coders into three categories:

\begin{itemize}[leftmargin=*]
\item \textbf{General-purpose LLMs:} Models such as GPT-4.1~\cite{openai2024gpt41}, Claude 3.5 Sonnet~\cite{anthropic2024claude35}, Gemini 2.5 Flash~\cite{geminiteam2025gemini25}, and Llama 4 Maverick~\cite{meta2024llama4} are evaluated for their ability to interpret diverse environments and generate coherent, well-structured code.
\item \textbf{Reasoning-enhanced LLMs:} These include MiniMax-M1~\cite{chen2025minimax}, DeepSeek-R1~\cite{guo2025deepseek}, O3-Mini~\cite{openai2025o3mini}, and Qwen3-235B~\cite{yang2025qwen3}, which emphasize explicit reasoning or planning abilities and may enhance performance on strategic or long-horizon tasks.
\item \textbf{Code-specialized LLMs:} Models such as Qwen-2.5-Coder-32B~\cite{hui2024qwen2}, Codestral 2501~\cite{mistral2024codestral}, Mercury-Coder~\cite{du2024mercury}, and Codex-Mini~\cite{openai2025codexmini} are optimized for code synthesis and are assessed for their adaptability to full agent implementation.
\end{itemize}

\begin{table*}[t!]
\caption{LLM coder tournament rankings and performance metrics across all game environments. Columns under \emph{Average Game-based Tournament Rankings} report each model’s average ranking based on TrueSkill across nine games. \emph{Code Quality \& Robustness} columns summarize development and runtime metrics: \textbf{A.R.} is the mean number of code revision iterations (max 3); \textbf{Part. (\%)} is the percent of agents successfully participating in tournament play;  \textbf{Win (\%)} indicates the overall percentage of matches won by each model, aggregated across all games and rounds; \textbf{Error (\%)} measures the incidence of timeouts and runtime errors during play; \textbf{Speed (s)} denotes the agent's average decision time per step (in seconds) across all games.}
\vspace{-2pt}
\label{tab:proxywar-main}
\centering
\renewcommand{\arraystretch}{1}
\resizebox{\linewidth}{!}{
\begin{tabular}{l|cccc|cccc|c|cc|ccc}
\toprule
\multirow{4}{*}{\textbf{Model}}
& \multicolumn{9}{c|}{\cellcolor{cyan!10}\textbf{Average Game-based Tournament Rankings \ (\(\downarrow\))}}
& \multicolumn{5}{c}{\cellcolor{pink!10}\textbf{Code Quality \& Robustness}} \\
\cmidrule(lr){2-10} \cmidrule(lr){11-15}
& \multicolumn{4}{c|}{\textbf{Single-Player}}
& \multicolumn{4}{c|}{\textbf{Two-Player}}
& \multicolumn{1}{c|}{\textbf{Multi-Player}}
& \multicolumn{2}{c|}{\textbf{Developing}}
& \multicolumn{3}{c}{\textbf{Runtime}} \\
\cmidrule(lr){2-5} \cmidrule(lr){6-9} \cmidrule(lr){10-10} \cmidrule(lr){11-12} \cmidrule(lr){13-15}
& Sudoku & Maze & 2048 & Hanoi
& TTT & Connect4 & Reversi & Snake
& Hold'em
& A.R. & Part. (\%)
& Win (\%) & Error (\%) & Speed (s) \\
\midrule
\rowcolor{orange!7}
\multicolumn{15}{l}{\textit{\textbf{General-purpose}}} \\
Qwen2.5-72B & 6.0 & 12.2 & 11.0 & 9.8 & 11.4 & 9.0 & 11.4 & 14.4 & 18.0 & 0.69 & 84.4 & 28.3 & 0.10 & 0.052 \\
Claude3.5-Sonnet & 10.2 & 11.0 & 9.4 & 4.4 & 8.2 & 12.6 & 11.0 & 7.8 & 7.0 & 0.31 & \cellcolor{purple!25}\textbf{100} & 21.0 & 0.21 & 0.294 \\
Phi4 & 7.2 & 10.2 & 8.4 & 12.2 & 5.0 & 14.5 & 14.0 & 9.6 & 17.4 & 0.91 & 84.4 & 27.6 & 3.40 & 0.034 \\
Gemini2.0-Flash & 10.2 & 9.6 & 8.6 & 10.2 & 8.8 & 9.4 & 13.8 & 11.6 & 7.8 & 0.22 & 95.6 & 28.3 & 0.21 & 0.024 \\
DeepSeekV3-0324 & 3.4 & 4.4 & 7.2 & 8.0 & 6.0 & 5.6 & 9.2 & 10.2 & 8.0 & 0.20 & \cellcolor{purple!25}\textbf{100} & 35.9 & \cellcolor{purple!25}\textbf{0.07} & 0.056 \\
Llama4-Maverick & 17.4 & 5.8 & 6.8 & 11.4 & 14.0 & 8.6 & 9.0 & 12.4 & 14.8 & 0.62 & 95.6 & 21.5 & \cellcolor{purple!25}\textbf{0.07} & 0.059 \\
GPT4.1-Mini & 13.2 & 3.4 & 9.8 & 7.2 & 14.0 & 6.6 & 5.4 & 7.0 & 6.2 & 0.29 & \cellcolor{purple!25}\textbf{100} & 34.5 & 1.17 & 0.122 \\
GPT4.1 & 14.8 & \cellcolor{purple!25}\textbf{3.2} & 7.8 & 7.2 & 10.4 & 4.0 & 6.6 & 6.6 & 6.4 & 0.31 & 100 & 31.3 & 0.21 & 0.301 \\
\midrule
\rowcolor{lime!13}
\multicolumn{15}{l}{\textit{\textbf{Reasoning-enhanced}}} \\
O3-Mini-20250131 & 4.6 & 5.4 & 4.6 & 6.4 & 12.8 & 7.0 & 7.2 & \cellcolor{purple!25}\textbf{4.0} & 9.0 & 0.07 & \cellcolor{purple!25}\textbf{100} & 38.8 & 0.14 & 0.016 \\
Qwen3-235B-0428 & 12.6 & 14.2 & 6.2 & 12.8 & 8.8 & 8.0 & 11.0 & 7.0 & 13.6 & 0.38 & 84.4 & 34.9 & 0.33 & 0.069 \\
DeepSeek-R1-0528 & \cellcolor{purple!25}\textbf{1.4} & 13.8 & 9.4 & 12.8 & 12.2 & 5.4 & 6.4 & 6.6 & \cellcolor{purple!25}\textbf{3.4} & 0.53 & 91.1 & \cellcolor{purple!25}\textbf{39.6} & 0.14 & 0.123 \\
Claude4-Sonnet & 3.0 & 6.2 & 9.8 & \cellcolor{purple!25}\textbf{3.4} & \cellcolor{purple!25}\textbf{3.6} & \cellcolor{purple!25}\textbf{2.6} & \cellcolor{purple!25}\textbf{3.6} & 7.0 & 7.4 & \cellcolor{purple!25}\textbf{0.18} & \cellcolor{purple!25}\textbf{100} & 31.2 & 0.14 & 0.211 \\
Gemini-2.5-Flash & 13.8 & 8.6 & \cellcolor{purple!25}\textbf{4.0} & 7.6 & 7.8 & 13.4 & 7.6 & 8.8 & 5.8 & 0.44 & 97.8 & 25.1 & 0.35 & 0.176 \\
Magistral-Small & 9.0 & 13.4 & 16.6 & 8.4 & 10.8 & 14.0 & 10.4 & 9.0 & 8.6 & 1.20 & 97.8 & 19.7 & 1.64 & 0.010 \\
\midrule
\rowcolor{red!8}
\multicolumn{15}{l}{\textit{\textbf{Code-specialized}}} \\
Qwen2.5-Coder & 8.4 & 11.6 & 17.2 & 14.2 & 9.2 & 17.4 & 7.6 & 18.0 & 6.0 & 0.98 & 75.6 & 13.4 & 0.98 & 0.049 \\
Codestral-2501 & 12.6 & 14.8 & 15.0 & 17.0 & 9.6 & 7.0 & 13.2 & 8.8 & 12.2 & 0.60 & 86.7 & 31.1 & 0.58 & 0.050 \\
Mercury-Coder & 8.4 & 16.8 & 13.4 & 18.0 & 8.8 & 16.0 & 18.0 & 16.0 & 12.8 & 0.67 & 77.8 & 18.7 & 0.16 & 0.003 \\
Codex-Mini & 16.8 & 6.8 & 7.2 & 3.8 & 9.8 & 9.0 & 6.2 & 7.0 & 8.8 & 0.31 & 97.8 & 31.2 & 0.14 & 0.035 \\
\bottomrule
\end{tabular}
}
\smallskip
\end{table*}

\subsection{Prompt Engineering and Code Generation}

\begin{lstlisting}[style=proxywarpython, caption={Minimal BaseAgent interface.}, label={lst:baseagent}]
class BaseAgent(ABC):
    def __init__(self, name: str):
        self.name = name

    @abstractmethod
    def select_action(self, observation: Any, action_mask: List[bool]) -> Optional[int]:
        pass
\end{lstlisting}

The prompt manager in \method systematically transforms each game specification into a comprehensive set of coding instructions, providing LLMs with all information required to generate robust and compatible agent code. Rather than relying on minimal or loosely structured prompts, \method employs a consistent template that integrates the following key elements:

\begin{enumerate}[leftmargin=*]
    \item \textbf{Task Framing and Objectives:} Each prompt begins with an explicit task description, emphasizing not only functional correctness but also the goal of achieving strong competitive performance. The LLM is instructed to create an agent capable of winning against diverse opponents, encouraging solutions that go beyond naive or baseline strategies.
    \item \textbf{Game Environment Specification:} The prompt includes a detailed description of the game rules, a formal specification of the observation and action spaces, and concrete examples of data formats. This ensures the generated code can correctly interpret the game environment and select valid actions.
    \item \textbf{Code Structure Constraints:} To ensure seamless integration with the evaluation framework, the prompt specifies required class inheritance (from \texttt{BaseAgent}), mandatory method signatures (such as \texttt{select\_action}), import statements, file structure, and error handling expectations. These constraints minimize common LLM code generation errors and promote consistency across all submissions. The essential interface required of every agent is illustrated in Listing~\ref{lst:baseagent}.
    \item \textbf{Strategic and Algorithmic Guidance:} In addition to implementation requirements, the prompt encourages algorithmic creativity, game-theoretic reasoning, efficiency considerations, and comprehensive code documentation. This guidance aims to stimulate more sophisticated and maintainable agent designs.
\end{enumerate}

\subsection{Testing Infrastructure}
\label{sec:testing_infrastructure}

The testing infrastructure in \method uses a hierarchical design to ensure generated agents are correct and robust before tournaments, filtering common failure modes at multiple levels:

\begin{enumerate}[leftmargin=*]
    \item \textbf{Code Structure \& Validation:} This foundational layer verifies that submitted code meets basic requirements: file existence and readability, Python syntax correctness using AST parsing, proper class structure inheriting from \texttt{BaseAgent}, and interface compliance with required methods like \texttt{select\_action}.
    \item \textbf{Basic Functionality:} Tests fundamental agent capabilities, including basic action selection behavior, action validation to ensure returned actions are legal and properly formatted, solution format verification for puzzle games, and basic puzzle interaction mechanics.
    \item \textbf{Game Interaction \& Logic:} Evaluates advanced game-specific behaviors through multi-move game interaction tests, puzzle-solving logic validation, maze navigation capabilities, and scenario-specific testing (e.g., all-in situations in poker).
    \item \textbf{Robustness \& Performance:} Tests agent resilience and efficiency through edge case handling (empty action masks, terminal states), error condition recovery, response time validation, and scalability under increased complexity.
\end{enumerate}

All test results are compiled into structured reports and fed back into the repair loop, enabling iterative code refinement and significantly improving the rate of successful agent deployment.

\subsection{Tournament Management Engine}

The tournament management engine in \method coordinates competition among generated coders across diverse games and experimental settings. The system is responsible for scheduling matches, isolating code execution, handling errors, and collecting evaluation metrics. Key components include:

\begin{enumerate}[leftmargin=*]
\item \textbf{Flexible Match Scheduling:} For two-player games, the engine uses round-robin tournaments, pairing each coder in both player roles to mitigate first-move advantage. Multiplayer games adopt randomized or Swiss-system scheduling for efficient ranking, while single-player games use standardized challenge sets for consistent evaluation.
\item \textbf{Execution Isolation and Resource Control:} Each agent runs in a separate process with strict resource limits, such as memory and per-move time constraints. File system and network access are restricted to prevent information leakage or external calls, and any violation results in forfeiture or penalty.
\item \textbf{Failure Detection and Handling:} The engine detects and records failures such as timeouts, exceptions, illegal actions, and repeated crashes, applying defined penalties in tournament scoring and rating.
\item \textbf{Performance and Reliability Monitoring:} During matches, the system logs metrics including decision time, resource consumption, move validity, and exception statistics, supporting post-hoc analysis and reproducible benchmarking.
\end{enumerate}

\section{Empirical Evaluation}

In this section, we present an empirical evaluation of ProxyWar across 18 state-of-the-art LLMs and 9 diverse game environments. Our experiments aim to answer the following research questions:

\begin{itemize}[leftmargin=*]
\item \textbf{RQ1 (Discriminative Power):} Does ProxyWar's competition-based evaluation provide better discrimination between models compared to traditional code generation metrics?
\item \textbf{RQ2 (Quality-Performance Relationship):} What is the relationship between static code quality metrics and competitive performance in game environments?
\item \textbf{RQ3 (Environmental Sensitivity):} How do different game characteristics (complexity, information structure, player count) affect model rankings and performance patterns?
\end{itemize}

\subsection{Experimental Setup}

Experiments ran on a workstation with an Intel i9-12900KF (3.19\,GHz), 64\,GB RAM, and an NVIDIA RTX A5000 (24\,GB). All code generation and tournaments ran on this single machine to avoid hardware variability. For each environment, we ran five rounds; in each round, every model generated a fresh agent and all agents played a full tournament under identical settings. Overall, each model participated in over 10{,}000 matches, providing robust estimates. We fixed random seeds for prompt sampling and environment initialization within each round to reduce uncontrolled randomness. Each game enforced a 45\,s per-decision timeout; agents exceeding it were disqualified for that round, while faster responses contributed to observed latency behavior.

\subsection{Model Discriminative Power (RQ1)}

Table~\ref{tab:proxywar-main} shows that ProxyWar exposes clear performance differences among models that appear similar by traditional metrics. For example, Qwen2.5-Coder and DeepSeek-R1 may achieve comparable pass@k on static benchmarks, but ProxyWar reveals a nearly threefold gap in tournament win rates (13.4\% vs. 39.6\%).

\noindent\textbf{Fine-grained Differentiation.} Some models achieve 100\% participation, meaning their code always passes all tests and would be marked as ``task-complete'' by conventional standards. However, ProxyWar shows that their operational characteristics can vary widely among these models. For example, DeepSeekV3 and Claude 3.5 Sonnet both reach perfect participation, yet DeepSeekV3 outperforms Claude 3.5 Sonnet in win rate by nearly 15 percentage points. ProxyWar also reveals distinct strengths between similar models: GPT-4.1 ranks higher than GPT-4.1-Mini in Maze, while GPT-4.1-Mini performs better in Reversi. These gaps would be largely invisible in purely pass/fail settings where both models simply \textit{solve} most benchmark tasks.

\noindent\textbf{Category-Level Insights.} Code-specialized models often underperform in competitive, context-rich scenarios. Their average win rate is 23.6\%, which is lower than both general-purpose models (28.6\%) and reasoning-enhanced models (31.6\%). This suggests that being optimized for code completion does not guarantee effective program synthesis or strategic reasoning. In complex environments that require broader context and multi-step planning, general models tend to excel. More broadly, ProxyWar highlights that once basic correctness is satisfied, factors such as decision quality under constraints, timeout incidence, and robustness against diverse opponents become key discriminators between models.

\begin{finding}
Competitive, game-based evaluation provides fine-grained differentiation among LLM coders, revealing substantial performance gaps in operational behavior that are invisible to conventional pass/fail metrics.
\end{finding}

\subsection{Code Quality Analysis (RQ2)}

Table~\ref{tab:proxywar-test-report} summarizes the results of hierarchical code testing and clarifies the relationship between static code quality metrics and competitive performance.

\noindent\textbf{Beyond Functional Correctness.} Most models achieve high Pass@1 rates, often exceeding 90\%. However, these results do not predict real-world performance in dynamic environments. For instance, O3-Mini achieves the highest Pass@1 (97.3\%), but only ranks mid-tier in tournament play. DeepSeek-R1, with a lower Pass@1 (87.9\%), consistently wins more matches. The weak correlation (Spearman's $\rho$ = 0.23) between Pass@1 and competitive ranking shows that traditional metrics miss key aspects of practical code effectiveness.

\begin{figure}[t]
  \centering
  \begin{subfigure}[t]{0.48\linewidth}
    \begin{lstlisting}[style=proxywarpython, basicstyle=\ttfamily\scriptsize]
# DeepseekR1 Generated
def select_action(obs):
  def backtrack(g):
    for r in range(9):
      for c in range(9):
        if g[r][c]==0:
          for v in opts(g,r,c):
            g[r][c]=v
            if backtrack(g): return True
            g[r][c]=0
          return False
    return True
  ...
  backtrack(grid)
    \end{lstlisting}
  \end{subfigure}
  \hfill
  \begin{subfigure}[t]{0.48\linewidth}
    \begin{lstlisting}[style=proxywarpython, basicstyle=\ttfamily\scriptsize]
# GPT4.1 Generated
def select_action(obs):
  empties.sort(key=lambda rc:len(cand(rc[0],rc[1])))
  def backtrack(i=0):
    if i==len(empties): return True
    r,c=empties[i]
    opts=cand(r,c)
    opts.sort(key=lcv)
    for v in opts:
      ...
      if backtrack(i+1): return True
      ...
    return False
  backtrack()
    \end{lstlisting}
  \end{subfigure}
  \caption{\textbf{Case study.}
  Left: DeepSeek-R1 generates a minimal, fast backtracking agent.
  Right: GPT-4.1 uses advanced heuristics (MRV/LCV), but the Python implementation is slower in practice.}
  \Description{Side-by-side code listings illustrating a case study of LLM-generated Sudoku agents. The left subfigure shows a Python backtracking-based agent generated by DeepSeek-R1 with a simple recursive search strategy. The right subfigure shows an agent generated by GPT-4.1 that applies additional heuristics such as minimum remaining values and least constraining value ordering. The figure contrasts different algorithmic strategies used by the two models for the same task.}
  \label{fig:sudoku-case}
\end{figure}

\noindent\textbf{Algorithmic Efficiency versus Practical Performance.} As illustrated in Figure~\ref{fig:sudoku-case}, competitive evaluation reveals unexpected gaps between theoretically optimal algorithms and practical performance. In Sudoku, DeepSeek-R1 generates a minimalist backtracking agent, while GPT-4.1 applies advanced search heuristics, with Minimum Remaining Value (MRV) and Least Constraining Value (LCV), in a more complex implementation. Despite its theoretical superiority, GPT-4.1’s code is nearly 28 $\times$ slower in Python due to higher interpretation overhead and complex bookkeeping, resulting in worse tournament outcomes. DeepSeek-R1’s concise design not only solves puzzles faster but also generalizes better to harder instances. This demonstrates that, practical efficiency and implementation overhead can outweigh algorithmic sophistication, and these differences only emerge clearly under competitive, dynamic evaluation.

\noindent\textbf{Debugging and Repair Ability.}  
Interactive, multi-stage code generation provides new insight into LLMs’ debugging skills. Some models, such as Claude3.5-Sonnet and DeepSeekV3, achieve perfect repair rates, reliably fixing initial bugs via self-correction, while others like Mercury-Coder almost never succeed. However, even perfect repair does not guarantee strong tournament performance, revealing the limits of current debugging capabilities.

\noindent\textbf{Runtime Stability and Computational Efficiency.}
Models with low runtime error rates, such as DeepSeekV3 and Llama4-Maverick, consistently outperform less stable competitors. Beyond correctness and stability, we observe dramatic disparities in computational efficiency: for the same agent task, the average step time (\emph{Speed (s)}) across models may differ by up to two orders of magnitude (e.g., Claude3.5-Sonnet at 0.294s vs. Magistral-Small at 0.010s per step). This demonstrates that even for functionally correct code, resource usage and latency can vary substantially, which may impact deployment in resource-constrained or latency-sensitive environments.

\begin{finding}
Traditional static metrics such as pass@k fail to predict performance in dynamic or competitive settings. Practical efficiency, debugging ability, runtime stability, and computational resource usage are all crucial but often overlooked dimensions of the operational characteristics of generated programs.
\end{finding}

\subsection{Environmental Sensitivity (RQ3)}

Model performance demonstrates pronounced variation across different game environments, highlighting both specialized strengths and inherent limitations.

\noindent\textbf{Game Complexity Effects.} In simple games such as Tic-Tac-Toe (state space $10^4$), models tend to converge, with rankings tightly clustered and a low standard deviation ($\sigma = 2.8$). However, as complexity increases, such as in Reversi ($10^{28}$ states), the performance gap widens significantly ($\sigma = 3.6$). Here, models like Claude4-Sonnet (rank 3.6) and GPT-4.1-Mini (rank 5.4) stand out for their strong position evaluation, while code-specialized models, for example Mercury-Coder (rank 18.0), consistently underperform.

\noindent\textbf{Impact of Information Structure.} Perfect information games favor models with effective search and planning, as seen with Claude4-Sonnet’s dominance. In contrast, games with imperfect information, such as Texas Hold’em, shift the advantage to models excelling at probabilistic reasoning and opponent modeling, with DeepSeek-R1 (rank 3.4) performing especially well. The weak correlation between model rankings in perfect and imperfect information games ($\rho = 0.28$) further underscores the distinct skill sets required.

\noindent\textbf{Algorithmic Paradigm Specialization.} Single-player puzzles reveal each model’s algorithmic tendencies. DeepSeek-R1 excels in constraint satisfaction tasks (Sudoku, rank 1.4), GPT-4.1 demonstrates strong graph search abilities (Maze, rank 3.2), while Gemini-2.5-Flash is most effective in probabilistic planning (2048, rank 4.0). These trends persist across repeated trials, suggesting that observed strengths reflect model biases rather than random fluctuation.

\noindent\textbf{Consistency Across Games.} The variance in rankings across all environments varies by model category. General-purpose models show moderate consistency, reasoning-enhanced models have slightly higher variance, and code-specialized models display the most inconsistency, suggesting their optimizations are often narrow and environment-dependent. Notably, Claude4-Sonnet is an exception, maintaining high consistency across all games, while O3-Mini consistently ranks among the top five in two-player games.

\begin{table}[t!]
\caption{
LLM coder testing report. \textbf{Pass@1} is the test pass rate of the initial code. \textbf{Repair} is the bug fix rate from initial to final submission. \textbf{Structure}, \textbf{Function}, \textbf{Logic}, and \textbf{Robustness} assess code quality at four hierarchical testing layers, as detailed in Section~\ref{sec:testing_infrastructure}.
}
\label{tab:proxywar-test-report}
\centering
\renewcommand{\arraystretch}{1}
\resizebox{\linewidth}{!}{
\begin{tabular}{l|cccccc}
\toprule
\textbf{Model}
& \textbf{Pass@1}
& \textbf{Repair}
& \textbf{Structure}
& \textbf{Function}
& \textbf{Logic}
& \textbf{Robustness} \\
\midrule
\rowcolor{orange!7}
\multicolumn{7}{l}{\textit{\textbf{General-purpose}}} \\
Qwen2.5-72B & 0.915 & 0.710 & 0.974 & 0.943 & 0.733 & 0.850 \\
Claude3.5-Sonnet & 0.948 & \cellcolor{purple!25}\textbf{1} & 0.983 & 0.943 & 0.933 & 0.913 \\
Phi4 & 0.844 & 0.789 & 0.965 & 0.871 & 0.533 & 0.738 \\
Gemini2.0-Flash & 0.959 & 0.533 & 0.991 & 0.943 & 0.967 & 0.938 \\
DeepSeekV3-0324 & 0.940 & \cellcolor{purple!25}\textbf{1} & \cellcolor{purple!25}\textbf{1} & 0.936 & 0.933 & 0.863 \\
Llama4-Maverick & 0.918 & 0.900 & 0.922 & 0.900 & 0.967 & 0.925 \\
GPT4.1-Mini & 0.951 & \cellcolor{purple!25}\textbf{1} & 0.991 & 0.950 & 0.967 & 0.888 \\
GPT4.1 & 0.926 & \cellcolor{purple!25}\textbf{1} & 0.983 & 0.907 & 0.900 & 0.888 \\
\midrule
\rowcolor{lime!13}
\multicolumn{7}{l}{\textit{\textbf{Reasoning-enhanced}}} \\
O3-Mini & \cellcolor{purple!25}\textbf{0.973} & \cellcolor{purple!25}\textbf{1} & \cellcolor{purple!25}\textbf{1} & 0.964 & 0.967 & \cellcolor{purple!25}\textbf{0.950} \\
Qwen3-235B & 0.929 & 0.769 & 0.887 & 0.957 & 0.933 & 0.938 \\
DeepSeek-R1 & 0.879 & 0.955 & 0.904 & 0.857 & 0.967 & 0.850 \\
Claude4-Sonnet & 0.934 & \cellcolor{purple!25}\textbf{1} & \cellcolor{purple!25}\textbf{1} & 0.921 & 0.833 & 0.900 \\
Gemini2.5-Flash & 0.932 & \cellcolor{purple!25}\textbf{1} & 0.991 & 0.936 & 0.867 & 0.863 \\
Magistral-Small & 0.811 & 0.986 & 0.817 & 0.879 & 0.733 & 0.713 \\
\midrule
\rowcolor{red!8}
\multicolumn{7}{l}{\textit{\textbf{Code-specialized}}} \\
Qwen2.5-Coder & 0.858 & 0.692 & 0.957 & 0.864 & 0.800 & 0.725 \\
Codestral-2501 & 0.890 & 0.750 & \cellcolor{purple!25}\textbf{1} & 0.907 & 0.700 & 0.775 \\
Mercury-Coder & 0.890 & 0 & \cellcolor{purple!25}\textbf{1} & 0.893 & 0.667 & 0.813 \\
Codex-Mini & 0.967 & \cellcolor{purple!25}\textbf{1} & 0.965 & \cellcolor{purple!25}\textbf{0.986} & \cellcolor{purple!25}\textbf{1} & 0.925 \\
\bottomrule
\end{tabular}
}
\end{table}

\begin{finding}
Environmental factors such as complexity, information structure, and task type induce large performance swings. No single model excels universally, and model strengths are often context-dependent.
\end{finding}

\subsection{Implications for Model Selection \& Practice}

Our results underscore the necessity for multi-dimensional, context-aware evaluation. Practitioners should consider both the target environment and the required robustness when selecting LLM coders. Optimization for static code completion alone is insufficient for deployment in dynamic or adversarial scenarios.

\begin{finding}
Model selection should align with the complexity and information structure of the target application. Game-based evaluation surfaces weaknesses and strengths that may be missed by conventional metrics, guiding more reliable deployment.
\end{finding}

\section{Discussion}

\subsection{Limitations and Future Directions}

While ProxyWar provides a powerful new lens for LLM code evaluation, several limitations remain. Game environments, though rich and diverse, may not capture the full complexity of real-world software development. Scaling evaluation to broader or industry-scale problems poses engineering and cost challenges. Future work may extend this approach to larger codebases and more realistic collaborative scenarios.

\subsection{Threats to Validity}

While \method aims to provide a rigorous and comprehensive framework for evaluating code generation, several threats to validity remain. First, all LLM-based coders exhibit inherent stochasticity: results may vary across repeated runs due to randomness in sampling and prompt interpretation. While we use fixed seeds and prompt templates to improve consistency, absolute reproducibility may not be guaranteed, especially for proprietary models. Second, the selection of game environments, though diverse, may not fully capture the complexity and variety of real-world programming tasks. Conclusions drawn from performance on these games may not generalize to all domains or software engineering problems. We therefore view our environment suite as a starting point rather than an exhaustive benchmark, and we encourage future work to add new, non-game environments. Third, evaluation relies on specific implementations and external dependencies, such as Python interpreter versions and third-party APIs. Changes in LLM APIs or model updates over time could affect the replicability of our results. Finally, while tournament outcomes and static code metrics provide multi-dimensional assessment, no evaluation can exhaustively capture all aspects of code quality, including long-term maintainability or human factors. Our testing infrastructure remains subject to the usual limitations of test-based evaluation: inadequate or unrepresentative test suites may overestimate capabilities. We partially mitigate this through layered tests (structure, function, logic, robustness) and runtime monitoring (timeouts, exceptions, illegal moves), but independent auditing of the released test suites will be essential. We encourage future work to expand the benchmark suite, test additional environments, and further investigate these sources of variability.

\section{Related Work}

\subsection{Code Generation Evaluation}

LLMs have rapidly advanced the state of code generation, with models such as Codex~\cite{li2022competition}, Code Llama~\cite{roziere2023code}, StarCoder~\cite{chen2021evaluating}, and Qwen-2.5-Coder~\cite{hui2024qwen2} achieving strong results on standard programming benchmarks. However, evaluating the real-world capabilities of these models remains an open challenge.

Static function-level benchmarks~\cite{li2022competition, austin2021program, hendrycks2021measuring} remain the primary tools for code generation evaluation. These benchmarks provide natural language prompts and assess correctness based on predefined test suites, reporting metrics such as pass@k. While effective for measuring functional correctness on small, isolated problems, these settings lack coverage of practical software engineering scenarios, such as class-level generation, dependency management, or multi-file coordination~\cite{du2024evaluating, jimenez2024swebench}. Moreover, they do not capture critical aspects such as code maintainability, readability, efficiency, or style~\cite{yu2024codereval}. To address these limitations, recent efforts have proposed multi-dimensional and more realistic benchmarks. ClassEval~\cite{du2024evaluating} moves towards class-level evaluation, revealing significant drops in model performance compared to function-level tasks. SWE-bench~\cite{jimenez2024swebench} evaluates models on real-world bug fixes and repository-level modifications, requiring context-aware reasoning and multi-file edits. CoderEval~\cite{yu2024codereval} and LiveCodeBench~\cite{jain2024livecodebench} broaden evaluation to pragmatic tasks and aim to mitigate dataset contamination. In addition, CodeBLEU~\cite{ren2020codebleu} augments text-based metrics with syntax and data-flow similarity, and human preference studies, such as CodeArena~\cite{du2025codearena}, introduce side-by-side solution comparisons along style and efficiency dimensions.

Nevertheless, most current approaches still evaluate code in isolation, neglecting the interactive, iterative, and competitive aspects of actual programming. Only recently have frameworks like CodeAgent~\cite{zhang2024codeagent} and EvalPlus~\cite{liu2024evaluating} begun to model iterative self-repair and test-driven development. There remains a significant gap in capturing the collaborative or adversarial settings that shape software development outcomes.

\subsection{Games and Competition-Based Assessment}

Games have long served as benchmarks for AI, from classic Elo and TrueSkill systems~\cite{minka2018trueskill} to recent breakthroughs in games like StarCraft II~\cite{vinyals2019grandmaster} and Dota 2~\cite{berner2019dota}. In code generation, competition-based evaluation has appeared in frameworks such as CodeContests~\cite{li2022competition} and CodeContests+~\cite{wang2025codecontests+}, but these largely focus on static, contest-style problem-solving. More interactive or adversarial frameworks, such as CodeArena~\cite{du2025codearena}, enable head-to-head model comparison, but often remain limited to specific problem types. Recently, game environments~\cite{wang2023voyager, chen2023gamegpt, wang2025large, chen2024can, peng2025large} have been widely adopted to evaluate advanced LLM capabilities beyond code generation. Frameworks such as Voyager~\cite{wang2023voyager}, GameGPT~\cite{chen2023gamegpt}, LVLM-Playground~\cite{wang2025large}, VARP~\cite{chen2024can} use complex or interactive games to assess general reasoning, open-ended learning, or multimodal understanding in LLM-driven agents. However, these efforts primarily target general intelligence, agent coordination, or perception, rather than systematic program synthesis or code quality evaluation. \method advances this line of research by embedding code generation and evaluation within diverse, competitive game environments, supporting automated repair, skill-based ranking, and comprehensive code analysis. Unlike prior work, \method unifies competitive, multi-agent evaluation with static code analysis to more fully capture the capabilities and limitations of modern LLM-based coders.

\section{Conclusion}

We present \method, a unified framework for evaluating LLM-based code generation through competitive, game-based tournaments. By moving beyond static function-level benchmarks, \method enables multi-dimensional assessment of code correctness, efficiency, and strategic behavior in dynamic, adversarial environments. Our approach systematically integrates code generation, testing, repair, and tournament management, providing more realistic and actionable insights into the capabilities of modern code generation models. Looking ahead, \method opens new avenues for research in program synthesis and AI-driven algorithm discovery. Future work may explore whether LLM coders can create truly novel strategies or outperform hand-crafted agents in complex games, as well as extend the framework to broader domains and more open-ended tasks. We hope ProxyWar will serve as a foundation for rigorous, reproducible, and creative evaluation of next-generation program synthesis systems.

\begin{acks}
This work was supported by the Australian Research Council under Discovery Project DP260102534.
\end{acks}

\bibliographystyle{ACM-Reference-Format}
\bibliography{ref}



\end{document}